\documentclass[article]{JHEP3}

\usepackage{amsmath,amssymb}

\newcommand{\be}{\begin{equation}}
\newcommand{\ee}{\end{equation}}
\newcommand{\bea}{\begin{eqnarray}}
\newcommand{\eea}{\end{eqnarray}}
\newcommand{\bean}{\begin{eqnarray*}}
\newcommand{\eean}{\end{eqnarray*}}

\preprint{{\small \texttt{}}}

\title{On black hole thermodynamics and the entropy function formalism }
\author{Xian-Hui Ge  and Fu-Wen Shu\footnote{XHG and FWS contribute to this paper equivalently.}
\\
\\
 Asia Pacific Center for Theoretical Physics, Pohang 790-784, Korea\\
\\
\email{E-mail:gexh@apctp.org}, \quad \email{fwshu@apctp.org} }
 \abstract{~From the black hole thermodynamics point of view, we show that the
entropy function $\mathbf{f}$ and the free energy $F$ are related
via
$\mathbf{f}=e_{I}q_{I}+\Omega_Hq_I{A_{\phi}^I}'-\left.\frac{\partial
F}{\partial r}\right. |_{r_{H}}$. Assuming the entropy function is
known for extremal black holes, we propose an approach to
calculate the entropy of non-extremal cases by slightly moving the
extremal black hole geometry from extremality. The  entropy of
non-extremal $D1D5$- and $D1D5p$-branes in the presence of higher
derivative corrections are computed as concrete examples. An
attempt has also been made to explain why the entropy function
method can calculate the corrected entropy without knowing the
exact form of black hole solution in higher derivative gravity
theories. }

\keywords{Black Holes, Black Holes in String Theory}

\begin{document}



\vfill

\eject


\section{Introduction}
The entropy function formalism proposed by Sen,  turns out to be a
very powerful method in computing extremal black hole entropy in the
presence of higher derivative gravity terms \cite{sen}. Usually, the
entropy function method is  composed of three steps. The first is to
write near horizon
 geometry of an extremal black hole in $n$ dimensions into $AdS_2\times
 S^{n-2}$ with constant radii $v_i$. The next step is to assume the coupled electric,
 magnetic fields, and scalar fields to be some constants $u_i$.
 Finally, introducing a function (a function of $v_i$ and $u_i$), which is the integral of Lagrangian
 density over the horizon $S^{n-2}$ and performing the Legendre
 transformation of this function with respect to electric field
 strengths  and extremizing it with respect to the scalars $v_i$ and
 $u_i$. The entropy function method extends the understanding of the
 attractor mechanism by showing that not only scalar fields but also
 every parameter of the near horizon geometry of an extremal black
 hole can be fixed by extremizing a function evaluated near the
 horizon.

 Actually, the entropy function formalism makes the computation of Wald's
 entropy formula \cite{wald}, a more general formalism for computing
  black hole entropy in the presence of higher derivative terms,
  much simpler. Even though the entropy function formalism is
  initially claimed to work for supersymmetric extremal black holes in supergravity theories,
  there are several attempts to apply the entropy function to non-supersymmetric extremal black
  holes\cite{Goldstein,Tripathy,Goldstein1,Kallosh,SSen,AE,CPTY,CYY,chonam,NTT,xgao,GLP,ANYY}
  and even non-extremal black holes
  \cite{Garousi,CP3}. However, several problems are required to be clarified
  on the entropy function formalism:\\
  1). It is amazing that only from the near horizon geometry, the
  entropy function formalism computes the entropy of an extremal
  black hole by means of a Legendre transformation. However, people
  usually use the full spacetime geometry to compute the black hole
  entropy. Therefore, how to relate the entropy function formalism
  with the traditional understanding of black hole thermodynamics?
  Fortunately, this problem has been discussed by several authors in
  Ref\cite{walper,oscar}. The authors established the connection between
  the entropy function method and the traditional Euclidean approach
  for black hole thermodynamics at zero temperature limit and found
  that the entropy function agrees with the zero temperature limit
  of the Euclidean action.\\
  2). In some of the literature, the entropy function method has been extended to
  non-extremal black holes. However, there are no attractor
  mechanisms for non-extremal solutions. How to grantee that the
  extension of the entropy function formalism is safe? This is the
  main purpose of this paper. We first discuss the relations between
black hole thermodynamics and the entropy function formalism and
then we propose a method to compute the entropy of near extremal
black holes when higher derivative gravity terms are taken into
account. The main idea of our method is that we slightly move the
black hole geometry from extremality where the extremal entropy
function is still valid. In this case, the near-extremal black hole
entropy and the extremal black hole entropy function approximately
satisfy a equation $TS'+T'S\approx \mathbf{f}_{ex}$. One can solve
this equation to obtain the near-extremal black hole entropy.\\
  3). The near horizon  metric
  $ds^2=v_1(-r^2dt^2+\frac{dr^2}{r^2})+v_2d\Omega^2_{n-2}$ was used
  in obtaining higher derivative corrections to extremal black hole
  entropy. It seems that one can calculate the   corrected entropy
  without knowing the exact metric of extremal black holes in
  higher derivative gravity theory. We will explain this point by
showing that the $AdS_2\times S^{n-2}$ geometry does not change in
the presence of curvature square terms.

 It is found in the present paper that the entropy function
$\mathbf{f}$ and the free energy $F$ are related by
$\mathbf{f}=e_{I}q_{I}+\Omega_Hq_I{A_{\phi}^I}'-\left.\frac{\partial
F}{\partial r}\right. |_{r_{H}}$.  Our results  imply that
applying the entropy function method to non-extremal black holes
might have difficulties, because the near horizon geometry of
non-extremal solutions in higher derivative gravity theory are not
always $AdS_2\times S^{n-2}$. As to the third point, Sen
speculated that \textit{in any general covariant theory of gravity
coupled to matter fields, the near horizon geometry of a
spherically symmetric extremal black hole in $n$ dimensions has
$SO(2,1)\times SO(D-1)$ isometry}\cite{sen3}. In four and five
dimensions, this postulate was proved in \cite{footenote,AY,AF}.
The near horizon geometry of non-extremal black holes in the
presence of higher derivative gravity terms might not have the
$SO(2,1)\times SO(D-1)$ isometry, and also there is no attractor
mechanism for non-extremal black holes. However, in order to
overcome the above difficulties, we focus on a class of black
holes which have extremal correspondences and whose entropy can be
obtained from the entropy function formalism. Once the behavior of
an extremal black hole is known by using the entropy function
formalism, we put some ``excess'' energy into the extremal black
hole and make it ``near-extremal'' with low but non-zero
temperature. With this non-strict approach, we can calculate the
entropy of non-extremal black holes including the higher
derivative corrections.

The paper is constructed as follows. In section 2 we establish the
connections between black hole thermodynamics and the entropy
function formalism for non-extremal black holes. In section 3 we
review the relations between the near horizon geometry of extremal
black holes and the higher derivative gravity corrections. In
section 4 and section 5, we  give two examples of how to extend the
entropy function to non-extremal black holes when higher derivative
corrections are taken into account. Section 6 contains concluding
remarks.
\section{Entropy function and Euclidean black hole thermodynamics}
To be general, we start from an $n$-dimensional universal metric
\begin{equation} \label{umetric} ds^2=g_{ij}dx^{i}dx^{j}. \end{equation} According to Sen's entropy
function method, the metric is required to deform into a metric of
extremal black holes with near horizon geometry $AdS_2\times S^2$,
namely \cite{sen}
\begin{equation}
\label{demetric}
ds^2=v_1\left(-r^2dt^2+\frac{dr^2}{r^2}\right)+v_2d\Omega_{n-2}^2.
\end{equation}
where $v_1$ and $v_2$ are assumed to be constants. We do not deform
the metric (\ref{umetric}) into near horizon geometry and the
advantage of the deformed metric (\ref{demetric}) will be discussed
later. Now we denote $f$ as function of the lagrangian density
$\sqrt{-\rm det\ g}\ \mathcal{L}$ on the horizon with the following
form \footnote{The function defined here is different from that of
Sen's definition in that $ f=2\pi T'f_{sen}$. }
\begin{equation}
 f=\int dx^3...dx^{n}\sqrt{-\rm det\ g}\ \mathcal{L},
\end{equation} where the lagrangian density may include gravitational fields, scalar fields,
gauge fields and covariant derivatives of these fields i.e.
$\mathcal{L}=\mathcal{L}^{tree}+\mathcal{L}^{corr}$,  and
$\{x^3...x^{n}\}$ are the angular coordinates. The Euclidean action
can be obtained by doing a Wick rotation $t\rightarrow i\tau$,
i.e.\footnote{More precisely, by doing a Wick rotation $t\rightarrow
i\tau$, function $f$ should be replaced by
$$f_E\equiv \int dx^3...dx^{n}\sqrt{\rm det\ g_E}\
\mathcal{L}(i\tau).$$ However, we do not need to distinguish $f$
from $f_E$ since they have the same value.}
\begin{equation}
I_{E}=\int d\tau dr f.
\end{equation}
For a stationary black hole, we have
\begin{equation}
\label{Euclaction} I_{E}=\beta \int  dr f=\beta F,
\end{equation}
where $\beta$ is the Euclidean time and $F$ is the \textit{free
energy} according to Hawking and Page\cite{HP}.

Now we would like to discuss the relationship  between the function
$f$ and Wald entropy formula. Consider  the lagrangian as an
$n$-form $\mathbf{L}(\psi)$, where $\psi=\{g_{ab},R_{abcd},\Phi_{s},
F^I_{ab}\}$ denotes the dynamical fields considered in this paper,
including the spacetime metric $g_{ab}$, the corresponding Riemann
tensor $R_{abcd}$, the scalar fields $\{\Phi_s$, $s=0,1,\cdots\}$,
and the $U(1)$ gauge fields $F^I_{ab}=\partial_a A_b^I-\partial_b
A_a^I$ with the corresponding potentials $\{A_a^I$, $I=1,\cdots\}$.
Under this definition, the variation of $\mathbf{L}$ is
\begin{equation}
\delta\mathbf{L}=\mathbf{E}_{\psi}\delta \psi +d \mathbf{\Theta},
\end{equation}
where $\mathbf{\Theta}$ is an $(n-1)$-form, which is called {\it
symplectic potential form}, $\mathbf{E}_{\psi}$ corresponds to the
equations of motion for the metric and other fields. Let $\xi$ be
any smooth vector field on the space-time manifold, then one can
define a {\it Noether current form} as
\begin{equation}
\label{Noethercurrent}
\mathbf{J}[\xi]=\mathbf{\Theta}(\psi,\mathcal{L}_{\xi} \psi)-\xi
\cdot \mathbf{L}.
\end{equation}
We will consider $\xi$ to be a killing vector vanishing on the
bifurcation horizon. Thus, $\mathcal{L}_{\xi} \psi=0$ and $
\mathbf{\Theta}(\psi,\mathcal{L}_{\xi} \psi)=0$. The fact that
$d\mathbf{J}[\xi]=0$ will be preserved when the equations of motion
are satisfied shows that a locally constructed $(n-2)$-form
$\mathbf{Q}[\xi]$ can be introduced and an ``on shell'' formula can
be obtained
\begin{equation}
 \mathbf{J}[\xi]=d\mathbf{Q}[\xi]\, \label{onshell}.
\end{equation}
Wald's analysis based on the first law of black hole thermodynamics
showed that for general stationary black holes, the black hole
entropy is a kind of Noether charge at horizon \cite{wald} and can
be expressed as
\begin{equation}
S_{BH}=2\pi \int _{\mathcal{H}} \mathbf{Q}[\xi]\, ,
\end{equation}
where $\xi$ represents the Killing field on the horizon, and
$\mathcal{H}$ is the bifurcation surface of the horizon. It should
be noted that the Killing vector field has been normalized to have
unit surface gravity. Integrating over a Cauchy surface
$\bf{\mathcal{C}}$  on Eq.(\ref{onshell}) and using
Eq.(\ref{Noethercurrent}), we find that
\begin{equation}
\label{interJ}
\int_{\bf{\mathcal{C}}}\mathbf{J}=-\int_{\bf{\mathcal{C}}}\xi \cdot
\mathbf{L}=\int_{\bf{\mathcal{C}}}d\mathbf{Q}[\xi]\,
\end{equation}
In an asymptotically flat spacetime, we have the interior boundary
at the horizon $\mathcal{H}$ and  the outer boundary at infinity
$\infty$. We obtain
\begin{equation}\label{noethcharge}
\int_{\bf{\mathcal{C}}}d\mathbf{Q}[\xi]=\int_{\infty}\mathbf{Q}-\int_{\mathcal{H}}\mathbf{Q},
\end{equation}where we have used the Stokes theorem.
The Euclidean action $I_{E}$ corresponds to \cite{wald}
\begin{equation}
\label{Eu} I_{E}=-\frac{1}{T}\left[\int_{\bf{\mathcal{C}}}\xi
\cdot \mathbf{L}+\int_{\infty}t\cdot \mathbf{B}\right],
\end{equation}
where $\mathbf{B}$ is an $(n-1)$-form defined as
$$
\delta \int_{\infty}t\cdot\mathbf{B}=\int_{\infty}t\cdot
\mathbf{\Theta}.
$$
From Eqs.(\ref{Euclaction}), (\ref{interJ}), (\ref{noethcharge})
and (\ref{Eu}), one finds that
\begin{equation}
 F+\int_{\infty}t\cdot \mathbf{B}=\int_{\infty}\mathbf{Q}-\int_{\mathcal{H}}\mathbf{Q}
\end{equation}
Note that the ``canonical energy''  $\mathcal{E}$ and the
``canonical angular momentum'' $\mathcal{J}$ are defined by
\cite{wald},
\begin{eqnarray}
&&\mathcal{E}=\int_{\infty}(\mathbf{Q}[t]-t\cdot \mathbf{B}),\\
&&\mathcal{J}=-\int_{\infty}\mathbf{Q}[\varphi].
\end{eqnarray}
 For an asymptotically flat black hole metric, one can choose
the Killing vector as
$\xi^{a}=t^a+\Omega_{H}^{\mu}\varphi^a_{(\mu)}$ with $\Omega_H$
the angular velocity of the horizon, then one obtains
\begin{eqnarray}
\label{conserve}
F=\mathcal{E}-\Omega_{H}^{\mu}\mathcal{J}_{\mu}-\int_{\mathcal{H}}\mathbf{Q}[\xi^{a}]
\end{eqnarray}
The variation of Eq.(\ref{conserve}) leads to
\begin{eqnarray}
\label{conserveE} \delta F=\delta
\mathcal{E}-\Omega_{H}^{\mu}\delta\mathcal{J}_{\mu}-\delta\int_{\mathcal{H}}\mathbf{Q}[\xi^{a}]
\end{eqnarray}
The Noether charge is composed of the contributions from the $U(1)$
gauge fields and gravitational fields, that is to say
\begin{eqnarray}
\mathbf{Q}=\mathbf{Q}^{\rm F}+\mathbf{Q}^{ g}+...
\end{eqnarray}
where
\begin{eqnarray}
&&\mathbf{Q}^F_{a_1\cdots a_{n-2}}=\frac{\partial
\mathcal{L}}{\partial F^{I}_{ab}}\xi^c A_c^{I}\mbox{{\boldmath
$\epsilon$}}_{aba_1\cdots a_{n-2}}\, ,\\
&& \mathbf{Q}^g_{a_1\cdots
a_{n-2}}=-\frac{\partial\mathcal{L}}{\partial
R_{abcd}}\nabla_{[c}\xi_{d]}\mbox{{\boldmath
$\epsilon$}}_{aba_1\cdots a_{n-2}}\, .
\end{eqnarray}
Now, following the work of Ref.\cite{cai}, we consider a stretched
region near the horizon ranged from $r_{H}$ to $r_{H}+\delta r$,
i.e.
\begin{equation}
\delta\int_{\mathcal{H}}\mathbf{Q}[\xi^{a}]=\int_{r_H}^{r_{H}+\delta
 r}\left(\mathbf{Q}^{\rm F}[\xi^{a}]+\mathbf{Q}^{g}[\xi^{a}]\right
 )
\end{equation}
Taking account of the Killing equation, we have
$\nabla_{[a}\xi_{b]}=2\kappa \mbox{{\boldmath $\epsilon$}}_{ab}$
(where $\kappa$ is the surface gravity of the hole), and the two
parts are found to be\cite{cai}
\begin{eqnarray}
\label{equation1} &&\int_{r_H}^{r_{H}+\delta
 r}\mathbf{Q}^g[\partial_t+
\Omega_{H}\partial_{\phi}] =\delta r\left[\kappa'E+\kappa
E'\right]_{r_H}+\mathcal{O}(\delta
r^2)\, ,\\
\label{equation2} &&\int_{r_H}^{r_{H}+\delta
 r}\mathbf{Q}^F[\partial_t+
\Omega_{H}\partial_{\phi}] =-(q_I e_I+\Omega_{H}q_I A'_{\phi})\delta
r +{\cal O}(\delta r^2).
\end{eqnarray}
where $E(r)$ is defined as
\begin{equation}
E(r) \equiv -\int_{\mathcal{H}}\frac{\partial\mathcal{L}}{\partial
R_{abcd}}\mbox{{\boldmath $\epsilon$}}_{ab}\mbox{{\boldmath
$\epsilon$}}_{cd}dx^{1}\cdots dx^{n-2}\, .
\end{equation}
The above formula is exactly the Wald formula for entropy without
the factor $2\pi$ \cite{wald}. According to the definition of
entropy function in (4.4) of \cite{cai}, we find $E(r)$ is related
to the entropy  by $2\pi E(r_H)=S$. $e_I$ and the $U(1)$
electrical-like charges in Eq. (\ref{equation2}) are defined to be
\begin{eqnarray}
e_I &\equiv& F_{rt}^I(r_H), \label{eq34}\\
A'^{I}_{\phi}&=&(\partial_{\phi})^a A^{I}_{a},\\
 q_I &\equiv & -\int_{r}\frac{1}{(n-2)!}\frac{\partial
\mathcal{L}}{\partial F^{I}_{ab}}\mbox{{\boldmath
$\epsilon$}}_{aba_1\cdots a_{n-2}}dx^{a_1}\wedge\cdots \wedge
dx^{a_{n-2}}\, \nonumber\\
&=&-\frac{\partial }{\partial
e_I}\int_{r_H}\frac{\mathcal{L}}{2(n-2)!}\mbox{{\boldmath
$\epsilon$}}^{ab}\mbox{{\boldmath $\epsilon$}}_{aba_1\cdots
a_{n-2}}dx^{a_1}\wedge\cdots \wedge dx^{a_{n-2}}=\frac{\partial
f(r_H)}{\partial e_I}\, ,
\end{eqnarray}
where we have written $F_{ab}^I(r_H)$ as $-e_I \mbox{{\boldmath
$\epsilon$}}_{ab}$. If the near horizon extension $r_{H}\rightarrow
r_{H}+\delta r$ is also done for the free energy, we find that
\begin{equation}
\label{equation3} \delta F=-\int_{r_{H}}^{r_{H}+\delta r}f
dr=-f(r_{H})\delta r+{\cal O}(\triangle r^2)
\end{equation}
Substituting Eqs. (\ref{equation1}), (\ref{equation2}) and
(\ref{equation3}) into Eq. (\ref{conserveE}), we obtain
\begin{equation}
\label{equation4}\left(f(r_{H})-q_{I}e_{I}-\Omega_{H}q_I{A^{I}
_{\phi}}'\right)\delta r=\delta
\mathcal{E}-\Omega_{H}^{\mu}\delta\mathcal{J}_{\mu}-(S\delta T+ T
\delta S).
\end{equation}
This formula describes the first law of black hole thermodynamics
written in terms of $F$. In fact, one can show that
$(q_{I}e_{I}+\Omega_{H}q_I{A^{I} _{\phi}}')\delta r$ is related to
$\Phi \delta Q$ where $\Phi=\xi^aA_a|_\mathcal{H}$ is the
electrostatic potential on the horizon of the hole and
$Q=\int\frac{\partial \mathcal{L}}{\partial
F^{I}_{ab}}\mbox{{\boldmath $\epsilon$}}_{aba_1\cdots a_{n-2}}$ is
the electric charge\cite{gao}. According to Sen's entropy function
definition, the entropy function is defined by
\begin{equation}
\mathbf{f}=-f(r_{H})+q_{I}e_{I}+\Omega_{H}q_I{A^{I}
_{\phi}}'.\end{equation} Finally, we come to the equation for the
entropy function
\begin{equation}
\label{eqef}\mathbf{f}\delta r=\delta
\mathcal{E}-\Omega_{H}^{\mu}\delta\mathcal{J}_{\mu}-(S\delta T+ T
\delta S).
\end{equation}
From the first law of black hole thermodynamics, we know that
$\delta\mathcal{E}=T\delta
S+\Omega_{H}^{\mu}\delta\mathcal{J}_{\mu}$, and
\begin{equation}
\label{free}\mathbf{f}\delta r=S\delta T.
\end{equation}
It becomes clear that the entropy function is closely related to the
free energy via
\begin{equation}
\label{result}
\mathbf{f}=e_{I}q_{I}+\Omega_Hq_I{A_{\phi}^I}'-\left.\frac{\partial
F}{\partial r}\right. |_{r_{H}}.
\end{equation}
 Note that the
discussions on the relations between Wald entropy formula and the
entropy function method in Ref.\cite{cai} only works for
spacetimes with vanishing "canonical energy" i.e. $\mathcal{E}=0$
(for example pure de Sitter space or pure anti-de Sitter space),
because the boundary condition at infinity is not included in
their discussions. We also emphasize here that Eq.(\ref{eqef})
describes the relations between the black hole thermodynamics and
the entropy function in an asymptotically  flat space. When we
extend our discussions to asymptotically anti-de Sitter (AdS) or
de Sitter space, we need impose corresponding boundary conditions
because the mass definition in asymptotically AdS spaces  is
different from that in asymptotically  flat spaces.

In the zero temperature limit, our results agree with the results
obtained in Ref.\cite{walper,oscar}. We know that $F=TI_{E}$, so
Eq.(\ref{result}) becomes
\begin{equation}
\label{result2} \mathbf{f}=e_Iq_I+\Omega_{H}q_I{A^{I}
_{\phi}}'-TI'_{E}-T'I_{E}.
\end{equation}
When $T\rightarrow 0$, we find that
\begin{equation}
\label{result3} \mathbf{f}_{ex}=e_Iq_I+\Omega_{H}q_I{A^{I}
_{\phi}}'-T'I_{E}=ST'.
\end{equation}
We can redefine $\tilde{e}_{I}=e_I/T'$,
$\mathbf{\tilde{f}}=\mathbf{f}_{ex}/T'$ and ${\tilde{A}}^{I} _{\phi}
={A^{I} _{\phi}}'/T'$, and obtain
\begin{equation}
\label{result4}
\mathbf{\tilde{f}}=\tilde{e}_Iq_I+\Omega_{H}q_I{\tilde{A}}^{I}
_{\phi}-I_{E}=S.
\end{equation}
The above equation was first obtained in Ref.\cite{walper,oscar}.
Comparing the thermodynamic result Eq.(\ref{result4}) with the Sen's
entropy function, i.e.
\begin{equation}
S=2\pi(e_iq_i-f_{sen}).
\end{equation}
We have (see also \cite{walper,oscar})
\begin{equation}
\tilde{e}_{I}=2\pi e_{i},~~~q_I=q_i,~~~I_{E}=2\pi f_{sen}.
\end{equation}
We emphasize that $I_{E}$ in Eq.(\ref{result4}) is defined in the
zero temperature limit,in particular, the free energy is also
evaluated in the zero temperature limit. In this sense, the
Euclidean action in Eq.(\ref{result4}) is quiet different from
Euclidean action for the non-extremal black holes \cite{dutta}.

\section{Near horizon geometry and higher derivative curvature terms}

 Now let us
interpret Sen's entropy function method from Eq.(\ref{eqef}).
Usually according to Sen's entropy function method, if we can
deform the near horizon  geometry of extremal black holes into
$AdS_2\times S^{n-2}$, the calculation of Wald's entropy formula
$S_{BH}=-2\pi\int d^{n-2}x
\sqrt{h}\frac{\partial\mathcal{L}}{\partial
R_{abcd}}\mbox{{\boldmath $\epsilon$}}_{ab}\mbox{{\boldmath
$\epsilon$}}_{cd}\,$ becomes very simple. Remarkably, even without
knowing the solutions in higher derivative gravity theory, one can
obtain the higher order corrections to entropy of extremal black
holes (such as extremal 3- and 4-charge black holes in string
theory \cite{CP1,ag}). Here we explain why Eq.(\ref{demetric})
makes sense in obtaining higher derivative entropy corrections.

We give a demonstration for AdS spaces together with curvature
squared corrections.  The general action of $n$-dimensional
$R^2$-gravity with cosmological constant and matter. The action is
given by
\begin{equation}
\label{r2action} I=\int d^{n}x \sqrt{-g}\left\{\frac{1}{16\pi
G_{n}}R-\Lambda+aR^2+bR_{\mu\nu}R^{\mu\nu}
+cR_{\mu\nu\rho\sigma}R^{\mu\nu\rho\sigma}+\mathcal{L}_{m}\right\},
\end{equation}
where $a$, $b$, $c$ are arbitrary small coefficients derived in
string theory, $\mathcal{L}_{m}$ is the lagrangian for the matter
fields, and the negative (positive) constant $\Lambda$ creates an
AdS (dS) space with radius  $ l^2=-\frac{(n-1)(n-2)}{16\pi G_{n}
\Lambda}$. By the variation over the metric tensor $g_{\mu\nu}$, we
obtain the equation of motion, i.e. the Einstein equation with $R^2$
corrections
\begin{eqnarray}
\label{einstein} && \frac{1}{8\pi
G_n}\left(R_{\mu\nu}-\frac{1}{2}g_{\mu\nu}R+8\pi
G_{n}g_{\mu\nu}\Lambda\right)=T^{matter}_{\mu\nu}
+a(\frac{1}{2}g_{\mu\nu}R^2-2RR_{\mu\nu}+2\nabla_{\mu}\nabla_{\nu}R-2g_{\mu\nu}\square
R) \nonumber\\
&&+b\left(\frac{1}{2}g_{\mu\nu}R_{\rho\sigma}R^{\rho\sigma}+2\nabla_{\mu}\nabla_{\alpha}R^{\alpha}_{\mu}-\square
R_{\mu\nu}-g_{\mu\nu}\square
R/2-2R^{\alpha}_{\mu}R_{\alpha\nu}\right)\nonumber\\
&&+c\left(\frac{1}{2}g_{\mu\nu}R_{\alpha\beta\rho\sigma}R^{\alpha\beta\rho\sigma}
-2R_{\mu\alpha\rho\sigma}R_{\nu}^{~\alpha\rho\sigma}-4\square
R_{\mu\nu}+2\nabla_{\mu}\nabla_{\nu}R+4R^{\alpha}_{\mu}R_{\alpha\nu}-4R^{\alpha\beta}R_{\mu\alpha\nu\beta}\right)
\end{eqnarray}
 Since the near horizon geometry of extremal black holes  always
 has the geometry of $AdS_2\times S^{n-2}$, we may consider a pure
 AdS space in the Poincare coordinate, where the AdS metric is considered as the near horizon
 limit of an extremal black hole and temperature of the AdS space is therefore zero.
 The unperturbed  AdS metric reads
\begin{equation}
\label{ads}
ds^2=\frac{1}{z^2l^2}\left(-dt^2+d\vec{x}^2\right)+\frac{l^2}{z^2}dz^2,
\end{equation} where $z=1/r$ with respect to (\ref{demetric}). We
choose the metric (\ref{ads}) which is the Poincare coordinate that
covers part of the manifold, because the near horizon geometry of
extremal black holes is always $AdS_2\times S^{n-2}$. Following
Ref.\cite{campanelli}, we can calculate the curvature square
corrections to metric (\ref{ads}) straightforward. The perturbed
metric turns out to be
\begin{eqnarray}
\label{pertads}
ds^2=&&\frac{1}{z^2l^2}\left[-\left(1+\frac{16(n-4)\pi
G_{n}}{(n-2)l^2}((n-1)n
a+(n-1)b+2c)\right)dt^2+d\vec{x}^2\right]\nonumber\\
&&+\frac{l^2}{z^2\left(1+\frac{16(n-4)\pi G_{n}}{(n-2)l^2}((n-1)n
a+(n-1)b+2c)\right)}dz^2.
\end{eqnarray}
From (\ref{pertads}), one can find that, at least to the first
order of $a$, $b$ and $c$, the $R^2$ corrections have no effect on
the geometry of AdS space except the AdS scale. Therefore it is
safe to compute the higher curvature corrections to extremal black
hole entropy by using the unperturbed metric, i.e. the near
horizon geometry (\ref{demetric}). Actually, this is not restrict
to extremal black holes with near horizon geometry $AdS_2\times
S^{n-2}$.  We found that for pure de Sitter space the entropy with
higher derivative corrections can also be obtained without knowing
the corrected metric\cite{shg}.

\section{Entropy  for non-extremal $D1D5$-branes with
$R^4$ corrections} 
The entropy function formalism works well for extremal black holes,
since the near horizon geometry of extremal (i.e. BPS) black hole
has the exact geometry of $AdS_2\times S^{n-2}$. To extend the
entropy function formalism to non-extremal black holes, we will move
away from the extremality slightly. We assume the temperature of
non-extremal black holes $T_{H}\ll 1$ and then the entropy
$S=-T\frac{\partial I_{E}}{\partial T}-I_{E}\approx -I_{E}$. From
Eq.(\ref{result2}) and Eq.(\ref{result3}), we have
\begin{equation}\label{nonextremal} TS'+T'S\approx
\mathbf{f}_{ex}.\end{equation} Therefore, once we know the entropy
function for extremal black holes, we can obtain the non-extremal
black hole entropy by considering small non-extremality.  To make
sure the above proposal works well for non-extremal black hole when
higher derivatives corrections are included, we first make a doubt
check on non-extremal $D1D5$-branes, which was recently calculated
in Ref.\cite{ghodsi}.
\subsection{Entropy function method }
 The type II  supergravity action in string frame read
as
 \be S_{II}=\frac{1}{16\pi G_{10}}\int d^{10}x\sqrt{-g}\bigg\{
e^{-2\phi}\left(R+4\partial_\mu\phi\partial^\mu\phi-\frac{1}{12}H^2_{(3)}\right)-\frac12\sum_n\frac{1}{n!}F_n^2\bigg\}
\,,\label{IIB} \ee where $\phi$ is the dilaton, $H_{(3)}$ is NS-NS
3-form field strength, and $F_{(n)}$ is the electric R-R n-form
field strength where $n=1,3,5$ for type IIB supergravity and $n=2,4$
for type IIA supergravity.
  The
near horizon geometry of non-extremal $D1d5$-branes, which is given
by the following line-elements
\begin{eqnarray}
\label{D1D5} ds^2=\frac{r^2}{\sqrt{Q_1
Q_5}}\left[-(1-\frac{r_0^2}{r^2})dt^2+dy^2\right]+\frac{\sqrt{Q_1
Q_5}}{r^2}(1-\frac{r_0^2}{r^2})^{-1}dr^2&&+\sqrt{Q_1Q_5}d\Omega_3^2 +\sqrt{\frac{Q_1}{Q_5}}\sum_{i=1}^{4}dz_i^2 \nonumber\\
e^{-2\phi}=\frac{Q_5}{Q_1}, ~~~F_{rty}=2\frac{r}{Q_1},~~~F_{rtyz_1
...z_4}=2\frac{r}{Q_5}
\end{eqnarray}
We can see in the following that even we start from extremal $D1D5$
systems, the non-extremal $D1D5$-branes entropy can be obtained by
using (\ref{nonextremal}). The extremal metric $D1D5$-branes systems
reads
\begin{eqnarray}
\label{D1d5} ds^2=\frac{r^2}{\sqrt{Q_1
Q_5}}\left[-dt^2+dy^2\right]+\frac{\sqrt{Q_1
Q_5}}{r^2}dr^2&&+\sqrt{Q_1Q_5}d\Omega_3^2
+\sqrt{\frac{Q_1}{Q_5}}\sum_{i=1}^{4}dz_i^2
\end{eqnarray}
In order to apply the entropy function formalism to the  extremal
$D1D5$-branes, one should deform the near horizon geometry  to the
most general form which is the product of the AdS-Schwarzschild and
$S^3 \times T^4$ space,
\begin{eqnarray}
\label{defD15} ds^2=&&v_1\left\{\frac{r^2}{\sqrt{Q_1 Q_5}}\left[-(
dt^2+dy^2\right]+\frac{\sqrt{Q_1 Q_5}}{r^2} dr^2\right\}
+v_2\left\{\sqrt{Q_1Q_5}d\Omega_3^2
+\sqrt{\frac{Q_1}{Q_5}}\sum_{i=1}^{4}dz_i^2 \right\}\nonumber\\
&& e^{-2\phi}=\frac{Q_5}{Q_1}u,
~~~F_{rty}=2\frac{rv_1^{\frac{3}{2}}}{Q_1 v_2^{\frac{7}{2}}}\equiv
e_1,~~~F_{rtyz_1
...z_4}=2\frac{r}{Q_5}v_1^{\frac{3}{2}}v_2^{\frac{1}{2}}\equiv e_2,
\end{eqnarray}
where $v_1$, $v_2$, $u$ are assumed to be constants. The function
$f$ is defined to be the integral of the Lagrangian density over the
horizon, so from (\ref{IIB}) and (\ref{defD15}) we can write
\begin{eqnarray}
f(v_1,v_2,u,e_1,e_2,r)&\equiv&\frac{1}{16\pi G_{10}}\int
dx^H\sqrt{-g}\cal L\cr &&\cr &=&\frac{V_1V_3V_4r}{16\pi
G_{10}}Q_1^{3/2}Q_5^{-1/2}v_1^{3/2}v_2^{7/2}\nonumber\\
&\times &\left(\frac{6uQ_5^{\frac12}(v_1-v_2)}{Q_1^{\frac32}v_1v_2}+
\frac{Q_1^\frac12Q_5^\frac12}{2v_1^3r^2}e_1^2+\frac{Q_5^{\frac52}}
{2Q_1^{\frac32}v_1^3v_2^4r^2}e_2^2\right)\,,
\end{eqnarray} where $V_1$ is
the volume of $S^1$,  $V_3$ is the volume of  the 3-sphere with unit
radius, and $V_4$ is the $T^4$ volume. The entropy function is then
written as
\begin{eqnarray}
\label{threefour}
 \mathbf{f}_{ex}(v_1,v_2,u)&\equiv & \left(e_i\frac{\partial f}{ \partial
e_i}-f\right)  =\frac{V_1V_3V_4}{16\pi
G_{10}}v_1^{3/2}v_2^{7/2}r\left(\frac{6u(v_2-v_1)}{v_1v_2}+\frac{2}{v_2^7}
+\frac{2}{v_2^3}\right)
\end{eqnarray}
Now, we use the entropy function equation $TS'+T'S\approx
\mathbf{f}_{ex}(v_1,v_2,u) $, where the surface gravity is given by
$\kappa=\sqrt{g^{rr}}\frac{d}{dr}\sqrt{-g_{tt}}=\frac{r}{\sqrt{Q_1Q_5}}$.
Solving the entropy function equation, we find that
\begin{equation}
\label{mathe} S=\frac{V_1V_3V_4}{16 G_{10}}\sqrt{Q_1
Q_5}v_1^{3/2}v_2^{7/2}r\left(\frac{6u(v_2-v_1)}{v_1v_2}+\frac{2}{v_2^7}
+\frac{2}{v_2^3}\right).
\end{equation}
Extremizing  $S$ with respect to $v_i$ and $u_i$, we obtain
\begin{eqnarray}
\frac{\partial S}{\partial v_1}=0, ~~~\frac{\partial S}{\partial
v_2}=0,~~~\frac{\partial S}{\partial u}=0,
\end{eqnarray}
with the following solutions
\begin{equation}
 v_1=v_2=1,
u=1.
 \end{equation}

 Substituting the above values back to
Eq.(\ref{mathe}), we finally obtain the entropy for non-extremal
$D1D5$ branes
\begin{equation}
 S_{BH} \left. \right|_{horizon}=\frac{V_1V_3V_4 r_0
 \sqrt{Q_1Q_5}}{4G_{10}},
 \end{equation} where $r_0$ is the event horizon radius. The result
 agrees with that of Ref. \cite{ghodsi}, but the method here is rather
 simpler.

 We can see in the below that even when the higher derivative terms are
 included, the entropy function equation still works well. When the next
 leading order  Lagrangian for type II theory is included, the
 action becomes \cite{zanon}
\begin{equation}
\label{eff1}
 S=\frac{1}{16\pi G_{10}}\int d^{10}x\,\sqrt{-g}\bigg\{ {\cal L}^{tree}+
e^{-2\phi}\left(\gamma W\right)\bigg\},
\end{equation}
where $\gamma=\frac18\zeta(3)(\alpha')^3$ and $W$ can be written in
terms of the Weyl tensors \begin{equation}
W=C^{hmnk}C_{pmnq}{C_h}^{rsp}{C^q}_{rsk}+\frac12
C^{hkmn}C_{pqmn}{C_h}^{rsp}{C^q}_{rsk}\,. \end{equation} For the
metric configuration (\ref{defD15}), the contribution of the above
higher derivative terms to the entropy function
\begin{eqnarray}
\label{threeten} \Delta \mathbf{f}(v_1,v_2,u)=-\gamma
u\frac{V_1V_3V_4\sqrt{Q_1Q_5}}{16\pi G_{10}}v_1^{\frac32}
v_2^{\frac72}r\left[\frac{3(35 v_2^4-20 v_1^3v_2+18 v_1^2v_2^2 -20
v_1v_2^3+35 v_1^4)}{32Q_1^2Q_5^2v_1^4v_2^4} \right]
\end{eqnarray}
Substituting (\ref{threefour}) and (\ref{threeten}) back into the
entropy function equation $TS'+T'S\approx
\mathbf{f}_{ex}(v_1,v_2,u)+ \Delta \mathbf{f}(v_1,v_2,u)$, we find
that
\begin{eqnarray}
\label{three11}  S |_{r=r_0}=&&\frac{V_1V_3V_4}{16 G_{10}}\sqrt{Q_1
Q_5}v_1^{3/2}v_2^{7/2}r_0\left(\frac{6u(v_2-v_1)}{v_1v_2}+\frac{2}{v_2^7}
+\frac{2}{v_2^3}\right. \nonumber\\ &&\left. -\gamma \sqrt{Q_1Q_5}
\frac{3(35 v_2^4-20 v_1^3v_2+18 v_1^2v_2^2 -20 v_1v_2^3+35
v_1^4)}{32Q_1^2Q_5^2v_1^4v_2^4}  \right).
\end{eqnarray}
The final result can be obtained by extremizing $\mathcal{E}$ with
respect to $v_1$, $v_2$ and $u$, that is to say,
\begin{eqnarray}
\frac{\partial S}{\partial v_1}=0, ~~~\frac{\partial S}{\partial
v_2}=0,~~~\frac{\partial S}{\partial u}=0.
\end{eqnarray}
The solutions yield
\begin{eqnarray}
v_1=1-\gamma\frac{51}{32(Q_1Q_5)^{\frac32}}\,,\,\,v_2=1-\gamma\frac{27}
{32(Q_1Q_5)^{\frac32}}\,,\,\,
u=1+\gamma\frac{33}{8(Q_1Q_5)^{\frac32}}\,.\label{sol1}
\end{eqnarray}
We finally obtain the entropy of non-extremal D1D5-branes under
higher derivative corrections
\begin{equation}
 S_{BH}=\frac{V_1V_3V_4r_0\sqrt{Q_1Q_5}}{4G_{10}}\left[1-\gamma\frac{9}{8(Q_1Q_5)^{3/2}}
+O(\gamma^2)\right]\,.\label{finalS}
\end{equation}
Again, we reproduce the result of Ref.\cite{ghodsi}. Now it is safe
to extend this method to more general conditions.
\subsection{Wald's formula for the $R^4$ term}
We can also calculate the $R^4$ correction to the entropy of
$D1D5$-branes by using the Wald expression, which is given by
\begin{equation}
S_{BH}=-2\pi \int_{\mathcal{H}}d^3x\sqrt{h}\frac{\partial
L}{\partial
R_{\mu\nu\rho\sigma}}\epsilon_{\mu\nu}\epsilon_{\rho\sigma},
\end{equation}
where $\epsilon_{\mu\nu}$ is the binormal to the bifurcation
surface, and $\epsilon_{\mu\nu}\epsilon^{\mu\nu}=-2$. The Wald
entropy for AdS-Schwarzschild has been calculated in \cite{dutta}
and  we will follow the method in \cite{dutta} to calculate the
entropy for $D1D5$ systems. One can choose
$\epsilon_{\mu\nu}=\xi_{\mu}\eta_{\nu}-\xi_{\nu}\eta_{\mu}$, where
$\xi=\frac{\partial}{\partial t}$ is the Killing vector field, and
$\eta=-g_{tt}^{-1}\frac{\partial}{\partial
t}-\frac{\partial}{\partial r}$ is the null vectors normal to the
bifurcation Killing horizon. Now, keeping in mind that
$e^{-2\phi}=\frac{Q_5}{Q_1}$, we can rewrite the $R^4$ corrections
to the Lagrangian,
\begin{equation}
\Delta L=\frac{\gamma Q_5}{16\pi G_{10}Q_1}W
\end{equation}
Therefore, the corrections to the entropy is given by
\begin{equation}
\label{waldentropy} \Delta S_{BH}=-\frac{\gamma Q_5}{8
G_{10}Q_1}\int_{\mathcal{H}}dx^{H}\sqrt{h}\frac{\partial W}{\partial
R_{\mu\nu\rho\sigma}}\epsilon_{\mu\nu}\epsilon_{\rho\sigma}=-\frac{\gamma
Q_5}{8 G_{10}Q_1}\int_{\mathcal{H}}dx^{H}\sqrt{h}\frac{\partial
W}{\partial C_{abcd}}\frac{\partial C_{abcd}}{\partial
R_{\mu\nu\rho\sigma}}\epsilon_{\mu\nu}\epsilon_{\rho\sigma}
\end{equation}
Using the unperturbed metric Eq.(\ref{D1D5}), one can obtain
\begin{equation}
\frac{\partial W}{\partial C_{abcd}}\frac{\partial
C_{abcd}}{\partial
R_{\mu\nu\rho\sigma}}\epsilon_{\mu\nu}\epsilon_{\rho\sigma}=\frac{9}{4}\frac{1}{(Q_1Q_5)^{\frac{3}{2}}}.
\end{equation}Note that the integral of Wald entropy is defined over
the horizon $H=S^1\times S^3 \times T^4$ and
$\sqrt{h}=Q_1^{\frac{3}{2}}Q_5^{-\frac{1}{2}}r_0$. Finally, we
obtain
\begin{equation}
\Delta S_{BH}=-\gamma \frac{9V_1V_3V_4r_0}{32(Q_1Q_5)}
\end{equation}
The total entropy becomes
\begin{equation}
S_{BH}=\frac{V_1V_3V_4r_0\sqrt{Q_1
Q_5}}{4G_{10}}\left(1-\gamma\frac{9}{8(Q_1Q_5)^{\frac{3}{2}}}\right),
\end{equation}which agrees with Eq.(\ref{finalS}).
\section{Entropy of near-extremal $D1D5p$ black holes}
In this section, we extend the method of the modified entropy
function to calculate near-extremal 3-charge black holes. In the
following of our calculation, we assume that the 3-charge black
holes is near extremal with small temperature so that we can use the
extremal black hole entropy function and  Eq.(\ref{nonextremal}) to
compute the entropy of near-extremal  $D1D5p$ black holes.
\subsection{10-dimensional case}
The entropy of extremal $D1D5p$ black holes in the presence of
higher derivative corrections can be  easily computed using Sen's
entropy function \cite{ag}. By releasing the extremal constraint
slightly, we hope to compute the non-extremal 3-charge black hole
approximately. The near horizon geometry of $D1D5p$ metric is
$M_3\times S^3\times T^4$ where $M_3$ is the deformed $AdS_3$
geometry.

The non-extremal black hole metric in $n=10$ is given as follows in
string frame
\begin{eqnarray}
ds^{2}_{10}&=&\left(f_{1}(r)f_{5}(r)\right)^{-\frac{1}{2}}[-dt^{2}+dz^{2}+K(r)(\cosh
              \sigma dt+\sinh \sigma dz)^2] \nonumber \\
           & &+\left(f_{1}(r)/f_{5}(r)\right)^{\frac{1}{2}}dx^{2}_{\|}+f_{1}(r)^{\frac{1}{2}}f_{5}(r)^{\frac{1}{2}}
           [\frac{dr^{2}}{1-K(r)}+r^2d\Omega^{2}_{3}],\nonumber \\
e^{-2\phi} &=&\frac{f_{5}}{f_{1}},
\end{eqnarray}
where
\begin{equation}
K(r)=\frac{r_{0}^{2}}{r^{2}},~~~f_{1}(r)\equiv1+\frac{Q_1}{r^2}=1+\frac{r_{0}^{2}\sinh^{2}\alpha}{r^{2}},
~~~f_{5}(r)\equiv
1+\frac{Q_5}{r^2}=1+\frac{r_{0}^{2}\sinh^{2}\chi}{r^{2}},
\end{equation}
The conserved charges carried by this hole are
\begin{eqnarray}
N_1=\frac{Vr^2_0 \rm sinh2\alpha}{2g_s\alpha'^3},~~~N_5=\frac{r^2_0
\rm sinh2\chi}{2g_s \alpha'},~~~N_p=\frac{R^2_{z}Vr^2_0 \rm
sinh2\sigma}{2g^2_s \alpha'^4},
\end{eqnarray}
where $V=R_5R_6R_7R_8$, $R_i$($i=5,6,7,8$) denote the radii of the
four coordinates in $x_{\|}$, and $R_{z}$ is the radius of the
compact dimension $z$, along which there is a momentum $P$. The
extremal limit can obtained by
\begin{equation}
\label{trans} r_0\rightarrow 0,~~~\alpha\rightarrow \infty,
~~~\chi\rightarrow \infty,~~~\sigma \rightarrow \infty,
 \end{equation} while holding $Q_1, Q_5$, and $Q_p=r^2_0 \rm sinh^2\sigma$
fixed.
 The near horizon geometry of extremal $D1D5p$ system
has the following form
\begin{eqnarray}
\label{horimetric}
ds^2&=&v_1\bigg(\frac{Q_p-r^2}{\sqrt{Q_1Q_5}}dt^2+\sqrt{Q_1Q_5}\frac{dr^2}{r^2}+\frac{Q_p+r^2}{\sqrt{Q_1Q_5}}dz^2-2\frac{Q_p}{\sqrt{Q_1Q_5}}
dtdz\bigg) \nonumber \\
           &+&v_2\bigg(\sqrt{Q_1Q_5}d\Omega_3^2+\sqrt{\frac{Q_1}{Q_5}}dx_i^2\bigg).
\end{eqnarray} When $Q_p=0$, the geometry is $AdS_3\times S^3\times
T^4$. The entropy function for extremal $D1D5p$ black holes can be
obtained from (\ref{IIB}) and (\ref{horimetric}) (see also
\cite{ag}),
\begin{equation}
\mathbf{f}_{ex} = \frac{4\pi^3RV_{T^4}}{16\pi
G_{10}}rQ_1^{\frac32}Q_5^{-\frac12}v_1^{\frac32}v_2^{\frac72}\bigg[\frac{6u_s
(v_2-v_1)}{(Q_1Q_5)^\frac12
v_1v_2}+\frac{Q_5^{\frac12}}{Q_1^{\frac32}}(\frac{2}{v_2^3}+\frac{2}{v_2^7})
\bigg],
\end{equation}
where the value of $v_1$, $v_2$, $u_s$ can be obtained by
extremizing the entropy function
\begin{equation}
v_1=v_2=u_s=1.
\end{equation}
 Substituting the above formula into the equation
\begin{equation}
TS'+T'S\approx \mathbf{f}_{ex},
\end{equation} where $T=\frac{1}{ 2\pi r_0 \rm \cosh  \alpha
\cosh  \chi \cosh  \sigma}$,  again we obtain
\begin{equation}
\label{centropy} S_{BH}=2\pi\mathcal{E}=\frac{2\pi RV r^3_0}{g^2_s
\alpha'^4}\rm \cosh \alpha \cosh  \chi \cosh \sigma.
\end{equation}
Actually, the entropy functions for extremal $D1D5p$ black holes at
$n=5$ and $n=10$ are identified with each other. So, we reproduce
the entropy for non-extremal $D1D5p$ black holes.
\subsection{$R^4$ corrections to  non-extremal 10-dimensional $D1D5p$ black hole entropy}
The $R^4$ corrections to extremal 3-charge black hole entropy was
obtained in \cite{ag}. In this subsection, we will extend the
previous results to non-extremal case. Usually there are several
forms of higher order of corrections in type II superstring theory.
we can consider the higher derivative corrections in \cite{zanon}
\begin{equation}
\begin{array}{ll}
\label{R4}
L_{corr}=\gamma e^{-2\phi}(L_{1}-2L_{2}+\lambda L_{3}), \\
\\
L_{1}=R^{hmnk}R_{pmnq}{R_{h}}^{rsp}{R^{q}}_{rsk}+\frac{1}{2}R^{hkmn}R_{pqmn}{R_{h}}^{rsp}{R^{q}}_{rsk},\\
\\
L_{2}=R^{hk}(\frac{1}{2}R_{hnpk}R^{msqn}{R_{msq}}^{p}+\frac{1}{4}R_{hpmn}{R_{k}}^{pqs}{R_{qs}}^{mn}+R_{hmnp}{R_{kqs}}^{p}R^{nqsm}),\\
\\
L_{3}=R(\frac{1}{4}R_{hpmn}R^{hpqs}{R_{qs}}^{mn}+R_{hmnp}{{R^{h}}_{qs}}^{p}R^{nqsm}),\\
\end{array}
\end{equation}
where $\gamma=\frac{1}{8}\zeta(3)\alpha'^{3}$ and $\lambda$ is a
parameter which signifies the ambiguity in the field redefinitions
of the metric. Now the corrected entropy function becomes
\begin{eqnarray}
  \mathbf{f}_{ex}+\Delta \mathbf{f} &=&\frac{4\pi^3RV_{T^4}}{16\pi
G_{10}}rQ_1^{\frac32}Q_5^{-\frac12}v_1^{\frac32}v_2^{\frac72}\bigg[\frac{6u_s
(v_2-v_1)}{(Q_1Q_5)^\frac12
v_1v_2}+\frac{Q_5^{\frac12}}{Q_1^{\frac32}}(\frac{2}{v_2^3}+\frac{2}{v_2^7})\nonumber\\
 & &
-6\gamma u_s\frac{7v_1^4+7v_2^4+6\lambda
Q_5^\frac12(v_1^4-v_2v_1^3+v_2^4-v_1v_2^3)}{v_1^4v_2^4Q_1^2Q_5^2}\bigg],
\end{eqnarray}
The solutions to the equations of motion of moduli by extremizing
the corrected entropy function are found to be
\begin{eqnarray}
v_1=1+\frac{63 \gamma}{11 (Q_1Q_5)^{3/2}}, ~~v_2=1-\frac{91
\gamma}{11(Q_1Q_5)^{3/2}}, ~~u_s=1+\frac{567
\gamma}{11(Q_1Q_5)^{3/2}}.
\end{eqnarray} Then we obtain the corrected entropy function
\begin{equation}
\mathbf{f}_{ex}+\Delta \mathbf{f}=\frac{4\pi^3RV_{T^4}}{16\pi
G_{10}}r(1-\frac{84 \gamma}{(Q_1Q_5)^{3/2}})
\end{equation}
Then the entropy is given by
\begin{equation}
S_{BH}=\frac{2\pi RV r^3_0}{g^2_s \alpha'^4}\rm \cosh \alpha \cosh
\chi \cosh \sigma (1-\frac{84 \gamma}{(Q_1Q_5)^{3/2}}).
\end{equation}
When $\kappa\rightarrow 0$, we return to the extremal case with
$R^4$ corrections.

\section{Conclusions}

In summary, we have proved that the entropy function is related to
the free energy via
$\mathbf{f}=e_{I}q_{I}+\Omega_Hq_I{A_{\phi}^I}'-\left.\frac{\partial
F}{\partial r}\right. |_{r_{H}}$. In this sense the entropy function
is nothing special but a transformed version of black hole
thermodynamics. We have found that the secret of the entropy
formalism relies on the near horizon geometry of the metric. While
the near horizon geometry of a extremal black hole might have
$SO(2,1)\times SO(D-1)$ isometry (for spherically symmetric black
holes ) or $SO(2,1)\times U(1)$ isometry (for rotating holes) in
higher derivative gravity theory is not generally proved, we have
pointed out that the higher derivative terms do not change the
geometry of AdS space in Poincare coordinate. We have also proposed
that when we slightly move the black hole geometry from extremality
the extremal entropy function is still valid. In this case, the
near-extremal black hole entropy and the extremal black hole entropy
function approximately satisfy a equation $TS'+T'S\approx
\mathbf{f}_{ex}$. One can solve this equation to obtain the
near-extremal black hole entropy.  As concrete examples, we
calculate the entropy of non-extremal $D1D5$- and $D1D5p$-branes in
the presence of higher derivative corrections. Although the above
method is not rigorously established, our computations agree with
the results obtained in previous literature.

\textbf{Acknowledgements}\\
The authors would like to thank G. W. Kang, S. P. Kim, M. I. Park,
P.  Zhang and S. Q. Wu for their helpful comments at different
stages of this work.

\bibliographystyle{plain}

\end{document}